# Pangloss: Fast Entity Linking in Noisy Text Environments

Michael Conover
Workday, Inc.
michael.conover@workday.com

Matthew Hayes
Workday, Inc.
matthew.hayes@workday.com

Scott Blackburn
Workday, Inc.
scott.m.blackburn@workday.com

Pete Skomoroch
Workday, Inc.
peter.skomoroch@workday.com

Sam Shah
Workday, Inc.
sam.shah@workday.com

## ABSTRACT

Entity linking is the task of mapping potentially ambiguous terms in text to their constituent entities in a knowledge base like Wikipedia. This is useful for organizing content, extracting structured data from textual documents, and in machine learning relevance applications like semantic search, knowledge graph construction, and question answering. Traditionally, this work has focused on text that has been well-formed, like news articles, but in common real world datasets such as messaging, resumes, or short-form social media, non-grammatical, loosely-structured text adds a new dimension to this problem.

This paper presents *Pangloss*, a production system for entity disambiguation on noisy text. Pangloss combines a probabilistic linear-time key phrase identification algorithm with a semantic similarity engine based on context-dependent document embeddings to achieve better than state-of-the-art results (>5% in F1) compared to other research or commercially available systems. In addition, Pangloss leverages a local embedded database with a tiered architecture to house its statistics and metadata, which allows rapid disambiguation in streaming contexts and on-device disambiguation in low-memory environments such as mobile phones.

**ACM Reference Format:**
Michael Conover, Matthew Hayes, Scott Blackburn, Pete Skomoroch, and Sam Shah. 2018. Pangloss: Fast Entity Linking in Noisy Text Environments. In *KDD '18: The 24th ACM SIGKDD International Conference on Knowledge Discovery & Data Mining, August 19–23, 2018, London, United Kingdom.* ACM, New York, NY, USA, Article X, 9 pages. https://doi.org/10.1145/3219819.3219899

## 1 INTRODUCTION

Named entity disambiguation, the problem of linking natural language text with pointers to knowledge base entries, constitutes an essential stepping stone on the path to complex reasoning tasks like question answering and fact extraction [1, 21]. Such structured metadata underpins some of the most visible applications of machine learning, from the knowledge cards that populate search experiences to the reasoning engines that power voice assistant technologies [36]. Even conventional commercial machine learning problems such as content recommendation, query understanding, and targeted advertising can benefit from the structure afforded by connecting raw text to knowledge base entity identifiers.

As a field of study, entity disambiguation is a well-researched problem, with much work focused on algorithmic enhancements tailored to well-formed documents such as news articles [19, 25, 34]. A growing body of research investigates the challenge of disambiguating loosely formatted text like instant messages and tweets [15, 33], while little has been written about the architectural patterns that enable entity linking systems to perform at scale.

This paper introduces *Pangloss*, a production system for state-of-the-art entity disambiguation on messy text. Pangloss is based on probabilistic tokenization and context-dependent document embeddings, an approach that achieves superior (>5%) results in F1 on standard benchmarks relative to other commercial and research entity linking services. On a benchmark of noisy workplace messaging text, Pangloss performs at least 15% better in F1 compared to other systems.

Additionally, Pangloss uses an embedded database with a tiered caching architecture for its metadata storage, a design decision that ensures data locality at inference time. This allows Pangloss to perform low latency disambiguation and on-device entity disambiguation in low memory environments like mobile phones. Our experiments show that Pangloss has around twice the entity linking throughput at an order of magnitude reduction in memory pressure compared to Stanford CoreNLP [23], a commonly used entity linking framework in industry and academia. Pangloss has been in production for a year, performing entity linking on hundreds of millions of documents daily from a multitude of industries and domains.

The key contributions of this paper are:

(1) The description of an entity linking system running in production on heterogeneous inputs including news articles, instant messages, source code repository commit logs, and rich text from collaborative workplace documentation platforms;
(2) The successful application of probabilistic tokenization to key phrase identification for named entity linking, an approach which addresses shortcomings of traditional tokenization algorithms when applied to poorly formatted documents;
(3) A semantic engine that computes multiple document embeddings at inference time, a technique that models the distinct themes present in text that spans multiple topics;
(4) An architectural pattern centered around an embedded database that leverages tiered caching to enable memory efficient, compute-local metadata storage;

(5) This results in state-of-the-art performance on industry standard entity linking benchmarks in addition to top marks on a manually-annotated dataset of workplace instant messages.

The remainder of the paper is structured as follows. In Section 2, we review related work. Section 3 introduces the problem more formally and outlines Pangloss's algorithmic approach. Section 4 describes the system architecture and tradeoffs, with Section 5 describing the results achieved. We conclude the paper in Section 6.

## 2 RELATED WORK

Scientists and engineers alike have long sought to systematize human expression with structures amenable to computational reasoning [3, 5, 11]. Named entity recognition [29], coreference resolution [38], and word sense disambiguation [28, 30] represent efforts to this effect at the document and corpus level, where named entity linking is concerned with connecting passages to standardized resolution targets known as knowledge bases [34]. The past decades have seen a multitude of efforts to construct definitive knowledge bases [2, 39], with Wikipedia emerging as an effective and comprehensive resolution target [25, 27].

Though significant progress has been made on linking well-formed documents such as news articles [13, 19, 20], the loose formatting and sparse content of short-form document like tweets and instant messages remains challenging [14–16, 32, 41]. Several approaches have been introduced to address these issues, including enhancements to tokenization and entity recognition, as well as advances in representation learning, which allow systems to model the meaning of both words and entities in a sophisticated and expressive way [8, 17].

Of particular importance to modeling the semantic structure of documents has been the advent of high dimensional word embeddings, most notably popularized by the word2vec family of algorithms [26]. Entity linking and knowledge base completion algorithms tackling the problem of embedding knowledge base entries and even relationships themselves have also shown promise [6, 9, 37], with some efforts jointly embedding entities and text, an approach we build upon in this study [7, 40].

This work advances the field of study by describing the set of algorithmic approaches and architectural decisions that underlie a fast, accurate named entity linking system running in a production capacity. Additionally, we outline several novel applications of existing technologies, such as probabilistic tokenization, to the challenges associated with the kind of loosely-formatted content that constitutes a major share of text found in the modern workplace. Finally, we introduce an approach to quantifying the semantic similarity between documents and knowledge base entities that uses multiple, context-sensitive document embeddings to produce fine-grained representations of distinct textual themes on a real-time basis.

## 3 METHODOLOGY
### 3.1 Problem Formulation

The task of named entity linking is to connect a substring, or *surface form*, in a passage of text to its corresponding entry in a knowledge base [34]. To achieve this, an entity linking system must overcome the dual challenges of synonymy and polysemy—that is, there are many ways to refer to a cup of coffee, while *Tesla* can mean both man and machine. To overcome these issues, entity linking systems must take into account the semantic context in which a phrase occurs to make a determination as to which knowledge base entity, if any, is associated with the substring. This problem is especially difficult for short and poorly-formatted text such as instant messaging. The following sections will establish the concepts we use to structure our approach to the problem.

*3.1.1 Knowledge Base.* Knowledge bases contain structured information about entities and the relationships among them, often in the form of subject-predicate-object triples, e.g. (`Half Dome, height, 2,307m`) [2]. Knowledge bases take many forms, from domain-specific research projects like the Gene Ontology to all-encompassing, multilingual knowledge repositories like Wikidata. Of particular interest to industrial practitioners is the fact that knowledge bases are ripe with structured metadata that can be used as both features and prediction targets for machine learning systems [3, 36]. For our purposes, because of its broad coverage and wide distribution, Wikipedia makes an excellent resolution target. This corpus allows us to resolve references to concrete entities such as *Pablo Picasso* and *enigma machine* in addition to abstract concepts such as *gradient descent* or *universal basic income*. We note that while this work focuses on a specific resolution target, many of the techniques we outline, specifically those relating to semantic understanding and performance optimization, should be generalizable to other knowledge bases such as those found in the fields of medicine, law, finance, and governance.

*3.1.2 Surface Form.* The literature on entity linking uses the term *surface form* to describe the phrases in a passage, such as those italicized above, which can be linked to a knowledge base [34]. Notionally, one can think of the surface form as a specific manifestation of the underlying entity, with many superficial forms used to refer to the same concept. Surface forms can consist of unigrams or n-gram tokens, such as *apophenia* and *lensatic compass*.

*3.1.3 Entities.* Entries in a knowledge base such as Wikipedia are highly diverse and cover standard categories such as people, organizations and locations in addition to abstract concepts. For convenience, this work uses the word *entity* to refer to any entry in the knowledge base. Under the scenario in which a surface form does not have a corresponding entry in the knowledge base, an entity linking system should *abstain* from making a link prediction.

### 3.2 Data Flow & Algorithm

The entity linking procedure outlined in this paper, which in its most succinct form connects a phrase from a document to a specific Wikipedia entry, unfolds in the sequence of steps outlined in Figure 1. Given a passage of text, identify a set of surface forms which may have corresponding entries in a knowledge base. For each surface form, identify the set of candidate knowledge base entries to which it may refer. Next, in a process known as disambiguation, employ a learning to rank model to order candidate entries in terms of confidence that the surface form refers to each entry. Finally, because the disambiguation step always yields a top-ranked candidate regardless of whether an appropriate entry actually exists, optionally abstain from predicting a link between the surface form and the top-ranked entity.

### 3.3 Datasets

We describe the datasets used throughout this paper.

*3.3.1 Wikipedia.* Wikipedia provides an excellent starting point as a source of knowledge base entries. High-traffic, open source, and richly curated, its contents detail countless aspects of the world in an approachable and informative way. Conveniently, the Wikimedia foundation provides regularly-updated mirrors of the underlying data, with the corpus used for this study including more than 7 million individual pages, approximately 4 million of which represent distinct entities. In addition to the textual contents of each page, the dense hyperlink structure of Wikipedia also provides a great degree of leverage for understanding the natural language context in which entities are mentioned.

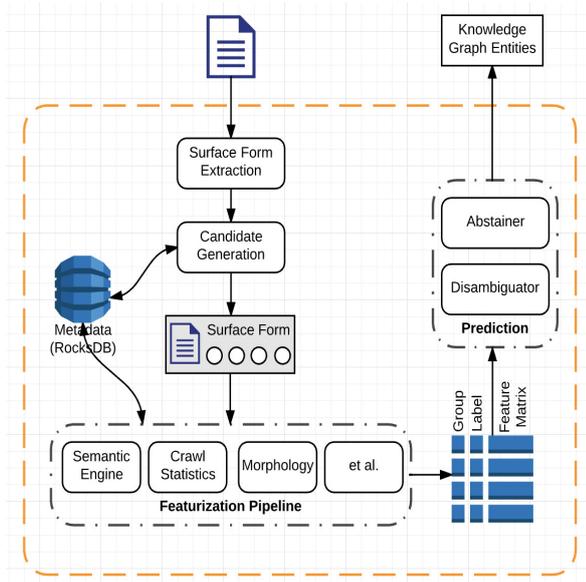

**Figure 1: Real-time entity linking by the Pangloss service proceeds in several stages.** A passage of text submitted to the system is algorithmically tokenized to identify substrings, or surface forms, which may have corresponding entries in Wikipedia (Section 3.4). For each surface form Pangloss calls a RocksDB key-value store to retrieve candidate entries (represented by circles) based on associations between hyperlink anchor text and Wikipedia URLs in Wikipedia and Common Crawl (Section 3.5). *(passage, surface form, candidate entry)* triples undergo several feature generation steps which include representational embedding using 300-dimensional vectors stored in RocksDB (Section 3.7). Candidate entries are scored using an XGBoost-based learning to rank model during the disambiguation phase, with the top-ranked candidate passed to a model which determines whether the system should abstain from forming the association (Section 3.8). Finally, the collection of linked entities is returned to the caller.

*3.3.2 Wikipedia Page Views.* The Wikimedia Foundation provides an auxilliary dataset, published hourly, describing the number of requests, both reads and edits, for individual pages on Wikipedia. Pangloss uses features derived from this dataset as an indicator of global popularity for knowledge base entries.

*3.3.3 LOD Wikilinks.* The LOD Wikilinks corpus is a coreference resolution corpus that contains over 40 million mentions of almost 3 million entities in the form of web-based hyperlinks to Wikipedia pages [35]. The anchor text for these hyperlinks describes a multitude of ways people refer to the concepts and entities in Wikipedia, and from this dataset we extract 10M documents consisting of more than 97M n-grams.

*3.3.4 Common Crawl.* The Common Crawl datasets represents a sample of web crawl data containing raw web page data, metadata and text extracts overseen by a 501(c)(3) nonprofit of the same name. Facilitating ease of access for industrial practitioners, the dataset is hosted for free on Amazon Web Services' Public Data Set repository in addition to academic hosts the world over.

As part of a batch Hadoop job run on a monthly basis we filter the Common Crawl data (~70TB) down to records which contain at least one hyperlink that points to English Wikipedia. This corpus has proven particularly valuable as a source of signal for associating tokens with knowledge base entries in the context of domain-specific, messy natural language.

*3.3.5 Workplace Chat Dataset.* In support of this research we also developed a manually-annotated dataset of workplace communication. Using instant messages from a popular IRC-like team chat application we identified threaded collections of messages corresponding to one hundred distinct conversations. Through a process outlined in Section 5.1, a team of annotators exhaustively identified the set of Wikipedia entries associated with the contents of the messages in the conversation.

## 3.4 Surface Form Extraction

The process for identifying phrases in a passage of text that correspond to knowledge base entries can have a significant impact on the performance of named entity linking systems [17]. This issue is especially acute for loosely-structured and poorly-formed documents such as workplace messaging, as traditional named entity recognition systems that rely on syntactical parses or word morphology do not yield sufficient recall to provide adequate linkage performance. Such a bottleneck at the top of the entity linking pipeline can have an outsized impact on downstream processes, and we observe that selecting the right key phrase identification algorithm, both in terms of computational efficiency and domain-appropriate functionality, is a critical component of deploying entity linking technology for commercial applications.

As such, we extend `segphrase`, an embarrassingly-parallelizable probabilistic phrasal segmentation algorithm [22] that uses an expectation maximization-based technique to partition sequences of words into disjoint subsequences corresponding to individual words and phrases. This approach balances the unlikeliness (as measured by phrasal pointwise mutual information) of a partitioning with an inferred quality score for individual phrases to produce sensible, domain-specific tokenizations in linear time.

As a source of raw count data for the `segphrase` algorithm, Pangloss utilizes phrases produced by passing multiple sliding n-gram windows over text from the corpora described in Section 3.3. To operationalize our domain-specific definition of a quality phrase, we hand labeled several thousand phrases, training data which was used as input to a binary classifier that took into account quality features from Liu et al. [22] in addition to predictors such as stop word incidence and the likelihood a phrase appeared as anchor text in Wikipedia. Using this classifier we derived "rectified" phrase-level frequency counts for Wikipedia n-grams based on the output of the phrasal segmentation step from Liu et al. [22]. Pangloss uses these rectified counts, alongside the quality scores outlined above, to compute phrase boundaries in novel input via dynamic programming-based segmentation at query time. We note here that it's straightforward to extend this framework to include additional features and sources of natural language data.

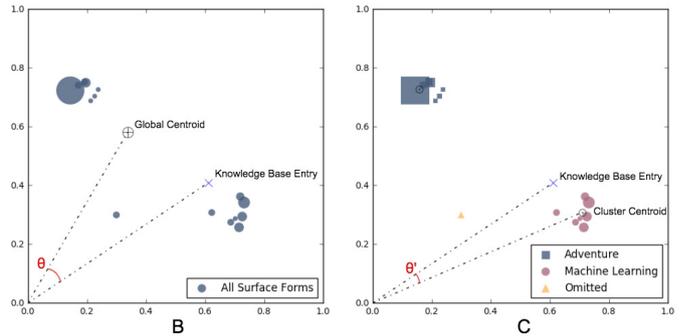

Figure 2: Workplace communication often interleaves distinct subjects in the course of a single conversational event (Figure 2A). Computing the TFIDF-weighted centroid of the embedding vectors representing all surface forms in a passage yields a coarse-grained passage vector that struggles to adequately represent the the entirety of the exchange (illustrated in Figure 2B). Consequently, the similarity between a candidate entity vector and the global centroid ($\theta$) will not reflect the fact that the candidate entity is highly similar to a specific conversational thread. Pangloss clusters surface form embeddings at query time, resulting in multiple document centroids corresponding to distinct themes. This mechanism allows us to produce multiple similarity features that describe the distribution of *(entity vector, cluster centroid)* similarity scores (illustrated in Figure 2C).

### 3.5 Candidate Generation

Given a *(surface form, text)* pair, one approach to identifying the correct entity with which to associate the surface form would be to train a multi-class classification model with a categorical cross-entropy loss function that assigns high scores to the correct entry in a knowledge base. In practice, as the reader may intuit, this does not work well owing to sparsity in the response variable and the volume of training data necessitated by such an arrangement. Instead, a common approach is to use the anchor text from links to knowledge base pages as a means to reduce the number of examples under consideration for any given surface form [34]. This recasts the problem in terms of ranking, and abstracts away the particulars of the individual surface forms and entities. To this end, Pangloss uses hyperlinks appearing in Wikipedia and in the Common Crawl dataset as a lookup table associating surface forms with candidate entities.

### 3.6 Joint Embedding

At the core of Pangloss is a mechanism for measuring the similarity between passages of text and Wikipedia entries. As input to this system, we compute representational embeddings of both knowledge base entries and surface forms side-by-side in the same vector space [7, 40]. To produce this embedding, we employ word2vec with the skip-gram objective function and negative sampling on the Wikipedia corpus described in Section 3.3.1.

To generate embeddings for multiword phrases, the system rewrites the text of each Wikipedia page using surface form extraction procedure described in Section 3.4. This procedure concatenates tokens from semantically meaningful n-grams to create individual tokens (e.g. *power_steering*) in the parsed documents for which unique embeddings can be learned in isolation of the constituent unigrams. To compute embeddings for knowledge base entries simultaneously with text, Pangloss utilizes internal Wikipedia hyperlinks, inserting a token representing a knowledge base entry (e.g. *uri:wiki/Wintermute*) following the anchor text linking to that article in an approach inspired by Yamada et al. [42]. In this way, word2vec is able to learn about the nature of knowledge base entries from the semantic context in which their inbound hyperlinks occur, and from this foundation we are able to embed both n-grams and documents in a common representational space.

### 3.7 Disambiguation Features

In the following sections we detail features the model relies upon to rank candidates as regards their likelihood of corresponding to the underlying sense of a surface form.

*3.7.1 Semantic Similarity.* Given a passage $p$ containing a surface form $s_i$, Pangloss needs to determine the similarity of a candidate knowledge base entry $e_{ij}$ to the passage $p$.

As a starting point, Pangloss computes a document representation consisting of the tfidf-weighted centroid of the embedding vectors corresponding to surface forms in $S$. This approach paints a very coarse picture of a document, and we observe that many passages address multiple subjects simultaneously, often in relation to one another, an effect that's especially pronounced in workplace content. Thus, to produce more fine-grained representations of the document, Pangloss generalizes the intuition of a document centroid to any set of surface forms, $S_k$, as follows:

$$V_S = \frac{\text{tf}_i \cdot \text{idf}_i \cdot v_i}{\sum_i \text{tf}_i} \quad (1)$$

In equation 1, $\text{tf}_i$ is the number of times surface form $s_i$ occurs in $S_k$, $v_i$ is the embedding vector representing surface form $i$, and $\text{idf}_i$ is calculated as the idf score of each term relative to the entire background corpus of documents, including Wikipedia and Common Crawl.

Using this formulation, Pangloss can represent subspaces of each passage with distinct centroids, capturing the subjects of a document with finer precision than the global mean. Using the intuition that the neighbors of a word are generally related, one set of relevant surface forms are those in a local window on either side of the surface form we are trying to disambiguate. For example, if *python* is the term in question, neighboring tokens such as *reptile* and *constrictor* would suggest the appropriate knowledge base entry is not the programming language. To determine the appropriate window size for this document vector Pangloss uses cross-validation and random search on the training set as a form of hyperparameter optimization.

Additionally, Pangloss groups together surface forms that are semantically related using HDBSCAN, a non-parametric clustering algorithm [24]. While allowing for the existence of a single cluster, this approach enables the algorithm to respect the meandering and multifaceted character of many texts. For example, consider a Monday morning workplace conversation in which the participants interleave small talk about the weekend with a casual discussion of the coming week's technical roadmap. That the embedding vector for a knowledge base entry is highly similar to the representation for one of two distinct clusters of tokens is more informative than finding it middlingly similar to a single monolithic document vector, as shown in Figure 2. It bears noting that for the cluster vectors, as with the local window vector, Pangloss uses cross-validation and random search to tune the minimum number of surface forms that must be present in a cluster for it to receive a corresponding embedding vector. We observe that this reduces variability associated with small cardinality clusters and contributes to improved model performance.

Thus, equipped with a collection of vectors representing distinct aspects of a document and a vector representing a knowledge base entry, we derive summary statistics describing the distribution of cosine similarity scores between each *(cluster, entity)* vector pair. For the vector representing terms in a local window around the target surface form, we find that both euclidean distance and cosine similarity contribute to improvements in model performance, as well as a feature denoting the number of surface forms contributing to the local window centroid. These features, taken together with an entity's cosine similarity to a single vector representing all surface forms in the document, constitute the semantic core of the Pangloss system.

*3.7.2 Anchor Text.* In the same way search engines are informed by the text associated with hyperlinks, so too can a named entity linking system learn from the anchor text associated with links into a knowledge base. For the purposes of this system, we consider hyperlinks to Wikipedia drawn from three principal sources: Wikipedia itself, the LOD Wikilinks corpus, and Common Crawl.

In the context of each of these datasets, for a surface form $s_i$, Pangloss counts the number of times the $s_i$ links to a specific entry $e_j$ in Wikipedia. As a control, Pangloss also computes the number of times the surface form occurs but is not linked. From this, the system can derive a variety of features characterizing the strength of the relationship between a surface form and a particular entry in the knowledge base. We define:

$$p_{ij} = \frac{a_{ij}}{\sum a_{i*}} \qquad (2)$$

to be equal to the probability that surface form $s_i$ links to entity $e_j$, where $a_{ij}$ is the number of times $s_i$ appears as anchor text in a hyperlink to $e_j$. Conceptually this is equivalent to $P(e_j|s_i)$, the conditional probability that a hyperlink with anchor text $s_i$ points to the knowledge base entry $e_j$.

Additionally, the system computes:

$$n_{ij} = \frac{a_{ij}}{\sum a_{i*} + a_{ix}} \qquad (3)$$

where $a_{ix}$ is the number of times $s_i$ appears but is not linked to the knowledge base. This feature helps to control for the incidence of terms which, when linked to the knowledge base, are linked with a high degree of certainty, but which in general do not warrant special attention. Consider the term *inside*—when *inside* appears as anchor text it often refers to a popular indie puzzle video game by the same name, however in general appearances of the word are not hyperlinked. Capturing this fact helps the model reason about its certainty regarding estimates of $p_{ij}$. Complementing these measures are several features which help to capture our uncertainty about both $n_{i*}$ and $p_{i*}$, including counts of the number of entries each surface form links to as well as the Shannon entropy of the distributions, as:

$$H(P_i) = \sum_j p_{ij} \cdot \log_2(p_{ij}) \qquad (4)$$

in the case of $P$. The set of features outlined in this section are computed for each corpus independently, such that Wikipedia, Common Crawl, and LOD Wikilinks each have a complement of distributions associated with their respective hyperlink profiles.

*3.7.3 Popularity.* Naturally, the popularity of public knowledge base entries should be indicative of the likelihood a surface form implies a given meaning. As a starting point, Pangloss leverages statistics derived from the Wikipedia dumps, some examples include: the number of inbound links, the total number of page views each entry receives, the number of pageviews via redirect pages, and the number of outbound links (total and unique). Additionally, Pangloss calculates for each of Wikipedia, LOD Wikilinks and Common Crawl the probability that any given hyperlink points to a specific entry. In addition, auxiliary features such as the Common Crawl-derived PageRank of each page would be appropriate to include as indicators of overall popularity.

*3.7.4 Morphology, String Similarity, et al.* Finally, we include in the model a variety of text-based signals that characterize the surface form, the Wikipedia page, and the relationship between the two. Some of these include the number of characters and words found on the Wikipedia page, the number of characters (upper- and lowercase) in the surface form, the number of whitespace separated words in the surface form, as well as the average word length and the maximum word length in the surface form. Additionally, Pangloss computes the probability that the surface form appears as hyperlink anchor text in each of Wikipedia, LOD Wikilinks and Common Crawl.

Simple string similarity measures comparing the surface form and the slug suffix of the Wikpedia URL corresponding to each entry (e.g. *Python_(Programming_Language)*) are also instructive. To this end, Pangloss calculates the Jaro-Winkler distance, which favors similarity in early characters; Damerau-Levenshtein distance, measuring the edit distance between two strings; and hash-based similarity measures.

## 3.8 Learning to Rank & Abstaining

*3.8.1 Training Data.* To train the disambiguation and abstain models described in the following section, we employed an iterative active-learning based approach to training data generation. These models require as input *(surface form, entity, passage)* tuples with labels indicating whether the entity is the correct sense of the surface form given the passage context. Initially we leveraged anchor text from Wikipedia hyperlinks in combination with the paragraph in which the hyperlink appeared as true positives. Negative examples are built from this set of positive examples by substituting other (incorrect) entity targets to efficiently produce a large volume of training data.

Improvements to this model correlated with performance gains on traditional corpora like news articles and Wikipedia but did not generalize to loosely-formatted content. To address this, we leveraged an intermediate version of the Pangloss system to identify entity predictions for this messy content about which the model exhibited maximal uncertainty. Subsequently, a team of manual annotators evaluated each marginal prediction, labeling records as either correctly or incorrectly linked based on the surrounding context. As before, we generated negative training data at scale by substituting correct entity associations with incorrect entity labels. For the abstain model, annotators were instructed to introduce one additional class—*None*—indicating that the surface form in question (e.g. *great news*), did not have a corresponding entry in the knowledge base.

*3.8.2 Disambiguation Model.* Given the training data outlined above, one could fit a binary classification model such as a support vector machine to predict whether a given *(surface form, text, knowledge base entry)* tuples represented a correct entity linkage. In practice, however, we find that this method fails to capture the full complexity of the problem [43], and that performance is improved by employing a pairwise preference algorithm that learns a model to determine which of two training examples should be ranked higher. For the purposes of this application, we find that pairwise XGBoost with cross-validated hyperparameters for gamma, learning rate, number of estimators, max depth, and minimum child weight yields the best relevance versus runtime performance.

*3.8.3 Abstain Model.* Ranking candidates is not the end of the story, however, as not every surface form identified in a passage of text should be linked to one of the candidates [34, 44]. For example, consider the case of project names in a workplace setting. A small-scale waterways environmental remediation project known as *Project Neptune* should not be linked to any of the Wikipedia pages associated with that phrase. As such, given the top-ranking candidate for a *(surface form, text)* pair, we train another model using the `Abstain` labeled data described above to predict whether a candidate can be meaningfully linked to an entry in Wikipedia. This formulation of the problem allows us to employ the feature extraction and modeling infrastructure developed for the disambiguation procedure while simply changing the response variable. Note that improvements to the abstain model can only improve precision, as recall is upper-bounded by the total number of pages identified as possible matches by the disambiguation algorithm.

## 4 IMPLEMENTATION

Pangloss depends on several types of metadata during training and prediction, including:

(1) `word2vec` vectors keyed by surface form and entity URL
(2) `segphrase` quality scores and segmentation probabilities keyed by phrase
(3) Statistics related to entities, keyed by entity URL
(4) Statistics related to surface forms, keyed by surface form

These word vectors and statistics are pre-computed offline using Apache Spark and Apache Hadoop and later fetched at inference time. In its raw form, this metadata is several gigabytes in size; compressed and loaded into memory as a monolithic asset requires considerable memory, prohibiting inference on cost effective cloud computing resources or small devices like mobile phones. This aspect of the system is problematic because the business requirements for systems such as Pangloss necessitate both high throughput and resource efficiency, where storing all the metadata in memory ensures high throughput but at the cost of resource efficiency. Notably, we observe that because the surface forms and phrases are extracted from natural language, they follow a roughly Zipfian distribution [31]. This means that as the popularity rank of a surface form or phrase decreases, the rate of metadata access decreases disproportionately. Devoting a large amount of memory to records that are infrequently accessed is wasteful but prohibits Pangloss from running in resource constrained environments.

Given this observation, Pangloss adopts a tiered architecture for its metadata storage. The first level of this architecture is an embedded database that stores the entirety of the metadata. Next, a set of key-level LRU caches, one for each data type, holds the most frequently accessed records in their deserialized form. Finally, a small, precomputed trie-based phrase filter eliminates a large number of unnecessary database lookups for non-existent keys.

Though there are many options for managing compute-local data access, Pangloss leverages RocksDB for its embedded database. RocksDB is a LevelDB-derived persistent key-value store database developed by Facebook and designed to be highly resource efficient without making undue performance sacrifices [12]. Other features that make RocksDB attractive are its fast bulk writing capabilities (which speeds construction of metadata assets, thus reducing model iteration time), high tunability (making it amenable to heterogenous computing environments), and ease of deployment (the database is a set of files in an archive).

A class serves as an abstraction for the metadata database, wrapping the RocksDB database and providing methods to fetch and deserialize each data type. Each method includes an LRU cache that stores return values in their deserialized form. This LRU cache enables the system to not only avoid database lookups but also the serialization and deserialization costs associated with these operations, as each lookup requires the key to be serialized and the value to be deserialized. Cache sizes are tunable to tradeoff linking throughput versus memory consumption.

Finally, the system utilizes a phrase prefix filter, which consists of a trie of all tokens that appear as the first token in at least one phrase stored in the database. This optimization resulted from the observation that during surface form extraction, a process that selects optimal phrase boundaries by exhaustively evaluating arbitrary partitions of input text, more than 90% of phrase lookups (*e.g.* for non-grammatical surface forms) yielded no result. Of the cases where the phrase wasn't found, 65% of the lookups could have been avoided by storing a cache of the initial phrase tokens. With this in mind, we assemble a trie and store it as side data alongside RocksDB. This asset—which consumes less than 2 MB on disk—results in a 30% reduction in surface form extraction time.

## 5 EVALUATION

Our evaluation answers the following questions:

(1) What is the quality of Pangloss's entity linking compared to other systems?
(2) What is the runtime performance of Pangloss?

We first compare Pangloss against other entity linking systems using standard benchmarks and a custom workplace benchmark. Second, we evaluate Pangloss's throughput and memory pressure by comparing it against a popular entity linking system used in industry and academia.

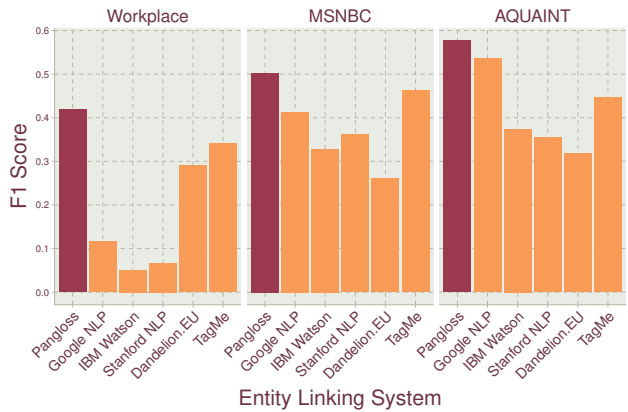

**Figure 3: F1-score of Pangloss entity linking performance relative to commercial and research systems on the MSNBC and AQUAINT benchmark news corpora in addition to a collection of manually-annotated conversations sampled from an enterprise chat application.**

## 5.1 Linkage Performance

In the following section, we establish Pangloss's state of the art named entity linking capabilities relative to commercial and research entity linking systems using two standard benchmark datasets and a manually-annotated dataset of conversations from workplace chatter.

The MSNBC dataset [10] consists of more than 600 manually linked surface forms drawn from 20 news articles covering 10 distinct subject areas. The AQUAINT dataset [27] consists of 50 Associated Press newswire stories linked to 449 Wikipedia entries by Mechanical Turk workers. The workplace dataset consists of a random sample of 100 threaded conversations reviewed by two annotators tasked with identifying the surface forms in each conversation that have corresponding Wikipedia pages (Section 3.3.5). We do not evaluate model performance on the popular TAC-KBP dataset owing to licensing restrictions governing the commercial use of the dataset.

Working independently, annotators were asked to exhaustively label 100 randomly selected message threads (1374 total judgements across 531 messages) from an enterprise chat application at a 1400 person social-good organization with links to Wikipedia pages associated utterances in the message. Annotators were instructed to prefer tight couplings between surface forms and Wikipedia entries; for example, on the basis of the tight coupling criteria an annotator would decline to form an association between the surface form *geophysicists* and the entry for *geophysics*, as the surface form refers to a group of people where the latter is a field of study. Additionally, annotators were instructed to avoid forming associations for routine conversational tokens such as *enjoy, thanks, later, great*, etc. Finally, annotators were instructed to annotate, where appropriate, complete surface forms rather than constituent tokens, for exampling creating a single association between a passage and the Wikipedia entry for *Google Calendar* rather than two associations, one for each of *Google* and *calendar*. We find a mean Jaccard coefficient between the sets of entities annotators assigned to each thread to be 0.56, and use as ground truth labels entities that both annotators identified for a given passage. We note that entities in the set difference were of generally high quality, and that a reasonable third party would likely conclude they were appropriate, and had simply been omitted due to slight differences in judgement. We also observe that considering the union of the two sets of annotators' judgements does not materially change the results reported herein.

We passed the text of each document to named entity linking APIs and tools, including the Google Cloud Natural Language, IBM Watson Natural Language Understanding, Stanford NLP [23] (version `2017-06-09`), Dandelion API, and TAGME [18]. The cloud providers were executed on May 22, 2018.

Comparing the response from each service to ground truth linked entities, we computed F1-scores (harmonic mean of precision and recall) characterizing the performance of each system as shown in Figure 3. Pangloss exhibits best of class performance on all three datasets, showing a 8.6% improvement over the next best system for MSNBC, a 7.6% increase for AQUAINT, and a 15.4% improvement relative to the second-ranked system on workplace conversations.

The differences between Pangloss and traditional named entity linking tools are most pronounced in the context of messaging-based workplace communication. To illustrate these differences more concretely consider the following exchange between three individuals coordinating around a visit to New York.

> **Case**: `i'll be visiting manhattan next week where i should go?`
> **Armitage**: `moma is amazing`
> **Molly**: `i can introduce you to a great machine learning team`

Exchanges such as these are characterized by short utterance lengths, inconsistent formatting and rapidly changing topical focus. Presented with this passage (actor names omitted) the IBM Watson Natural Language Understanding API is able to correctly link *Manhattan*, but does not link any of the other concepts mentioned in the passage. In contrast, owing to tokenization and disambiguation modules optimized for loosely-formatted input, Pangloss is able to correctly identify *Manhattan*, as well as the *Museum of Modern Art*, *machine learning* and the concept of a *team*. We observe that many natural language understanding frameworks, often built on top of traditional named entity recognition and syntax parsing tools, are highly sensitive to the morphology and structure of documents, limiting their ability to generalize to out of band content. This example, while succinct, captures the essence of how Pangloss differentiates itself from other named entity linking tools.

Notably, Pangloss' performance is significantly influenced by the initial tokenization of an input passage. For example, if the tokenization algorithm were to incorrectly break *machine* and *learning* into two distinct tokens the algorithm would fail to link the document to the knowledge base entry for *machine learning*.

Likewise, we observe that highly polysemous word can be difficult to link correctly in passages with limited semantic context. For example, consider the sentence *"Among technology startups, they have a very strong **network**."* Here, the semantic context suggested by the word *technology* could lead the algorithm to conclude the surface form refers to a computer network rather than a network of social connections.

## 5.2 Runtime Performance

The demands of production systems require fast and resource efficient algorithm performance [4]. To evaluate this aspect of the Pangloss system, we prepared a benchmark dataset of a consistent 0.01% random sample of Wikipedia pages where all markup was removed so that systems could operate on plain text. For the experiment, we used a dedicated machine with 20 physical cores (Xeon E5-2666 v3 Haswell processors at 2.9 GHz and supporting boosting up to 3.5 GHz) and 60 GB of RAM running Linux 3.19. We ran each analysis single-threaded though we see proportional performance as we scale to more threads. We compare the performance of Pangloss against Stanford CoreNLP version 2017-06-09 [23], an NLP system widely used in industry, academia, and government. For CoreNLP, we set the minimum set of annotator switches to enable entity linking and fix the maximum Java heap size to 8 GB.

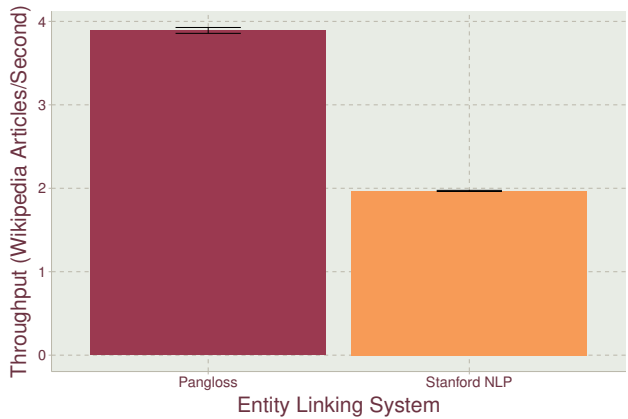

**Figure 4: A comparison of single-threaded throughput between Pangloss and Stanford CoreNLP for the Wikipedia pages entity linking benchmark. On average, Pangloss completed 3.9 Wikipedia pages/second and Stanford CoreNLP completed 2.0 Wikipedia pages/second. We exclude the considerable start up time of Stanford CoreNLP.**

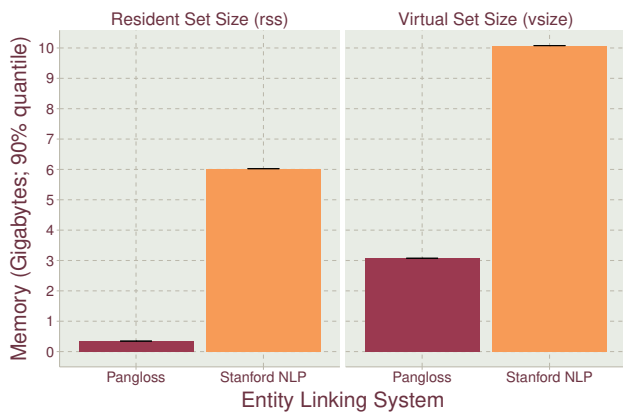

**Figure 5: The 90% quantile resident memory (rss) and virtual memory (vsize) consumed by Pangloss and Stanford CoreNLP for the Wikipedia pages entity linking benchmark. Pangloss uses 6% of the resident memory and 31% of the virtual memory of Stanford CoreNLP. We see consistent and similar results at any quantile level.**

Pangloss has almost twice the entity linking throughput at about 6% of resident memory of Stanford CoreNLP. Figure 4 shows a comparison of steady-state throughput of Pangloss vs Stanford CoreNLP for 3 runs of the Wikipedia page benchmark. We ignore the startup costs of Stanford CoreNLP in this experiment. On average, Pangloss links 3.9 Wikipedia articles per second while Stanford CoreNLP links 2.0 Wikipedia articles per second.

Figure 5 shows the memory consumption, resident set (rss) and virtual set (vsize), on this same benchmark. We sample the memory consumption every second on the Wikipedia page benchmark and compute the 90% quantile. Pangloss uses 6% of resident memory (0.345 GB vs 6.02 GB) and 31% of virtual memory (3.07 GB vs 10.1 GB) compared to Stanford CoreNLP. We see consistent and similar results throughout the run and at any quantile level. Stanford CoreNLP loads its metadata into resident memory for fast lookups; Pangloss is able to achieve superior performance with considerably less memory pressure through the use of its tiered metadata storage model.

## 6 CONCLUSION

Named entity linking draws together research threads from many distinct domains in natural language processing. Putting these ideas into practice for commercial applications demands a careful balance of algorithmic sophistication and strategic architectural choices. Our work in this area indicates that high quality disambiguation requires the careful tuning of multiple interrelated sub-modules, most notably the tokenization and semantic similarity apparatus. We find that for commercial applications, text pre-processing, fine-grained feature development, and parameter tuning constitute the difference between a good and a great system. Moreover, while pure research efforts can afford to pay little mind to the issues of latency and memory efficiency, deploying these technologies in a production environment absolutely necessitates resource conscious design. All told, this paper presents a novel approach to the problem of named entity linking, and outlines in detail the methodological and engineering decisions for a machine learning system used at scale.

## REFERENCES

[1] J. Andreas, M. Rohrbach, T. Darrell, and D. Klein. Learning to compose neural networks for question answering. *arXiv preprint arXiv:1601.01705*, 2016.
[2] S. Auer, C. Bizer, G. Kobilarov, J. Lehmann, R. Cyganiak, and Z. Ives. Dbpedia: A nucleus for a web of open data. In *The Semantic Web*, pages 722–735. Springer, 2007.
[3] T. Berners-Lee, J. Hendler, and O. Lassila. The semantic web. *Scientific American*, 284(5):34–43, 2001.
[4] R. Blanco, G. Ottaviano, and E. Meij. Fast and space-efficient entity linking for queries. In *Proceedings of the Eighth ACM International Conference on Web Search and Data Mining*, pages 179–188. ACM, 2015.
[5] D. M. Blei, A. Y. Ng, and M. I. Jordan. Latent dirichlet allocation. *Journal of Machine Learning Research*, 3(Jan):993–1022, 2003.
[6] A. Bordes, J. Weston, R. Collobert, Y. Bengio, et al. Learning structured embeddings of knowledge bases. In *AAAI*, volume 6, page 6, 2011.
[7] A. Bordes, X. Glorot, J. Weston, and Y. Bengio. Joint learning of words and meaning representations for open-text semantic parsing. In *Artificial Intelligence and Statistics*, pages 127–135, 2012.
[8] C. Cherry and H. Guo. The unreasonable effectiveness of word representations for Twitter named entity recognition. In *Proceedings of the 2015 Conference of the North American Chapter of the Association for Computational Linguistics: Human Language Technologies*, pages 735–745, 2015.
[9] K. Clark and C. D. Manning. Improving coreference resolution by learning entity-level distributed representations. *arXiv preprint*

*arXiv:1606.01323*, 2016.
[10] S. Cucerzan. Large-scale named entity disambiguation based on Wikipedia data. In *Proceedings of the 2007 Joint Conference on Empirical Methods in Natural Language Processing and Computational Natural Language Learning (EMNLP-CoNLL)*, 2007.
[11] S. Deerwester, S. T. Dumais, G. W. Furnas, T. K. Landauer, and R. Harshman. Indexing by latent semantic analysis. *Journal of the American society for information science*, 41(6):391, 1990.
[12] S. Dong, M. Callaghan, L. Galanis, D. Borthakur, T. Savor, and M. Strum. Optimizing space amplification in RocksDB. In *CIDR*, 2017.
[13] M. Dredze, P. McNamee, D. Rao, A. Gerber, and T. Finin. Entity disambiguation for knowledge base population. In *Proceedings of the 23rd International Conference on Computational Linguistics*, pages 277–285. Association for Computational Linguistics, 2010.
[14] Y. Fang and M.-W. Chang. Entity linking on microblogs with spatial and temporal signals. *Transactions of the Association for Computational Linguistics*, 2:259–272, 2014.
[15] S. Guo, M.-W. Chang, and E. Kiciman. To link or not to link? A study on end-to-end tweet entity linking. In *Proceedings of the 2013 Conference of the North American Chapter of the Association for Computational Linguistics: Human Language Technologies*, pages 1020–1030, 2013.
[16] Y. Guo, B. Qin, T. Liu, and S. Li. Microblog entity linking by leveraging extra posts. In *Proceedings of the 2013 Conference on Empirical Methods in Natural Language Processing*, pages 863–868, 2013.
[17] B. Hachey, W. Radford, J. Nothman, M. Honnibal, and J. R. Curran. Evaluating entity linking with Wikipedia. *Artificial Intelligence*, 194: 130–150, 2013.
[18] F. Hasibi, K. Balog, and S. E. Bratsberg. On the reproducibility of the TAGME entity linking system. In *Proceedings of 38th European Conference on Information Retrieval*, ECIR '16, pages 436–449. Springer, 2016.
[19] J. Hoffart, M. A. Yosef, I. Bordino, H. Fürstenau, M. Pinkal, M. Spaniol, B. Taneva, S. Thater, and G. Weikum. Robust disambiguation of named entities in text. In *Proceedings of the Conference on Empirical Methods in Natural Language Processing*, pages 782–792. Association for Computational Linguistics, 2011.
[20] S. Kulkarni, A. Singh, G. Ramakrishnan, and S. Chakrabarti. Collective annotation of Wikipedia entities in web text. In *Proceedings of the 15th ACM SIGKDD International Conference on Knowledge Discovery and Data Mining*, pages 457–466. ACM, 2009.
[21] A. Kumar, O. Irsoy, P. Ondruska, M. Iyyer, J. Bradbury, I. Gulrajani, V. Zhong, R. Paulus, and R. Socher. Ask me anything: Dynamic memory networks for natural language processing. In *International Conference on Machine Learning*, pages 1378–1387, 2016.
[22] J. Liu, J. Shang, C. Wang, X. Ren, and J. Han. Mining quality phrases from massive text corpora. In *Proceedings of the 2015 ACM SIGMOD International Conference on Management of Data*, pages 1729–1744. ACM, 2015.
[23] C. D. Manning, M. Surdeanu, J. Bauer, J. Finkel, S. J. Bethard, and D. McClosky. The Stanford CoreNLP natural language processing toolkit. In *Association for Computational Linguistics (ACL) System Demonstrations*, pages 55–60, 2014.
[24] L. McInnes, J. Healy, and S. Astels. HDBSCAN: Hierarchical density based clustering. *The Journal of Open Source Software*, 2(11):205, 2017.
[25] R. Mihalcea and A. Csomai. Wikify!: Linking documents to encyclopedic knowledge. In *Proceedings of the 16th ACM Conference on Information and Knowledge Management*, pages 233–242. ACM, 2007.
[26] T. Mikolov, I. Sutskever, K. Chen, G. S. Corrado, and J. Dean. Distributed representations of words and phrases and their compositionality. In *Advances in Neural Information Processing Systems*, pages 3111–3119, 2013.
[27] D. Milne and I. H. Witten. Learning to link with Wikipedia. In *Proceedings of the 17th ACM Conference on Information and Knowledge Management*, pages 509–518. ACM, 2008.
[28] A. Moro, A. Raganato, and R. Navigli. Entity linking meets word sense disambiguation: A unified approach. *Transactions of the Association for Computational Linguistics*, 2:231–244, 2014.
[29] D. Nadeau and S. Sekine. A survey of named entity recognition and classification. *Lingvisticae Investigationes*, 30(1):3–26, 2007.
[30] R. Navigli. Word sense disambiguation: A survey. *ACM Computing Surveys (CSUR)*, 41(2):10, 2009.
[31] D. M. Powers. Applications and explanations of Zipf's law. In *Proceedings of the joint conferences on new methods in language processing and computational natural language learning*, pages 151–160. Association for Computational Linguistics, 1998.
[32] A. Ritter, S. Clark, O. Etzioni, et al. Named entity recognition in tweets: an experimental study. In *Proceedings of the conference on empirical methods in natural language processing*, pages 1524–1534. Association for Computational Linguistics, 2011.
[33] W. Shen, J. Wang, P. Luo, and M. Wang. Linking named entities in tweets with knowledge base via user interest modeling. In *Proceedings of the 19th ACM SIGKDD International Conference on Knowledge Discovery and Data Mining*, pages 68–76. ACM, 2013.
[34] W. Shen, J. Wang, and J. Han. Entity linking with a knowledge base: Issues, techniques, and solutions. *IEEE Transactions on Knowledge and Data Engineering*, 27(2):443–460, 2015.
[35] S. Singh, A. Subramanya, F. Pereira, and A. McCallum. Wikilinks: A large-scale cross-document coreference corpus labeled via links to Wikipedia. *University of Massachusetts, Amherst, Tech. Rep. UM-CS-2012*, 15, 2012.
[36] A. Singhal. Introducing the knowledge graph: things, not strings. *Official google blog*, 2012.
[37] R. Socher, D. Chen, C. D. Manning, and A. Ng. Reasoning with neural tensor networks for knowledge base completion. In *Advances in Neural Information Processing Systems*, pages 926–934, 2013.
[38] W. M. Soon, H. T. Ng, and D. C. Y. Lim. A machine learning approach to coreference resolution of noun phrases. *Computational Linguistics*, 27(4):521–544, 2001.
[39] F. M. Suchanek, G. Kasneci, and G. Weikum. Yago: A core of semantic knowledge. In *Proceedings of the 16th International Conference on World Wide Web*, pages 697–706. ACM, 2007.
[40] Z. Wang, J. Zhang, J. Feng, and Z. Chen. Knowledge graph and text jointly embedding. In *Proceedings of the 2014 Conference on Empirical Methods in Natural Language Processing (EMNLP)*, pages 1591–1601, 2014.
[41] I. Yamada, H. Takeda, and Y. Takefuji. Enhancing named entity recognition in Twitter messages using entity linking. In *Proceedings of the Workshop on Noisy User-generated Text*, pages 136–140, 2015.
[42] I. Yamada, H. Shindo, H. Takeda, and Y. Takefuji. Joint learning of the embedding of words and entities for named entity disambiguation. *arXiv preprint arXiv:1601.01343*, 2016.
[43] Z. Zheng, F. Li, M. Huang, and X. Zhu. Learning to link entities with knowledge base. In *Human Language Technologies: The 2010 Annual Conference of the North American Chapter of the Association for Computational Linguistics*, pages 483–491. Association for Computational Linguistics, 2010.
[44] S. Zwicklbauer, C. Seifert, and M. Granitzer. Robust and collective entity disambiguation through semantic embeddings. In *Proceedings of the 39th International ACM SIGIR conference on Research and Development in Information Retrieval*, pages 425–434. ACM, 2016.